\documentclass[11pt,a4paper]{article}

\usepackage[utf8]{inputenc}
\usepackage[margin=1in]{geometry}
\usepackage{amsmath,amssymb}
\usepackage{siunitx}
\usepackage{graphicx}
\usepackage{authblk}
\usepackage{cite}
\usepackage{xcolor}
\usepackage{array}    
\usepackage{hyperref}
\usepackage{mathrsfs}

\hypersetup{
    colorlinks=true,
    linkcolor=blue,
    filecolor=magenta,
    urlcolor=cyan,
    citecolor=red,
}

\title{\textbf{On the Infeasibility of Low-Energy Warp Drive via Metamaterial Gravitational Coupling}}
\author{Jos\'{e} Rodal}
\affil{Rodal Consulting, Cary, NC 27513, USA \\ \href{mailto:jrodal@alum.mit.edu}{jrodal@alum.mit.edu}}

\date{\today}

\begin{document}
\maketitle

\begin{abstract}
Recent ``low-energy’’ warp-drive concepts propose replacing the constant gravitational coupling \(\kappa_{0}\) with a spatially varying scalar field \(\kappa(x)\) set by an engineered metamaterial’s electromagnetic response. We show that the idea fails on both theoretical and experimental grounds. A prescribed, non-dynamical \(\kappa(x)\) in the field equation \(G^{\mu\nu} = \kappa(x)T^{\mu\nu}\) clashes with the contracted Bianchi identity, \(\nabla_{\mu} G^{\mu\nu} \equiv 0\), forcing \(\nabla_{\mu}T^{\mu\nu} \neq 0\) and thus violating local energy–momentum conservation. Making \(\kappa(x)\) dynamical yields a scalar–tensor theory in which the scalar mediates a new long-range force that breaks the strong equivalence principle; Solar-System and pulsar-timing experiments already restrict \(|\gamma-1|\lesssim10^{-5}\), excluding any technologically useful coupling. Junction-condition analysis further shows that any interface where \(\kappa\) changes demands a \(\delta\)-function layer of stress–energy, while even steep continuous profiles are ruled out by torsion-balance, lunar-laser, and spacecraft Doppler measurements. Hence the concept is untenable: a non-dynamical scheme violates conservation laws, and its scalar-tensor completion is falsified by existing data.
\end{abstract}

\section{Introduction}
It has been speculated that one could engineer a material-dependent and hence position-dependent gravitational coupling,
\begin{equation}
  \kappa(x) \;=\; \mathscr{S}(x)\,\kappa_0
\end{equation}
with exotic ``metamaterials'' to mitigate the negative-energy demands of warp-drive metrics such as Alcubierre’s~\cite{Alcubierre1994,SarfattiAPS2025,Sarfatti2022Acad,Taylor2024}. Since any such device is finite, its internal $\kappa_{internal}$ must transition to the ambient value $\kappa_{ambient}$ at its interface. This requires a boundary layer where $\kappa(x)$ changes rapidly, creating a concentration of stress-energy whose magnitude depends on the specific model and must be carefully evaluated. We will demonstrate that this concept faces severe theoretical and experimental obstacles making it non-viable under current physical principles. 

\medskip
\noindent\textbf{Dynamical vs.\ prescribed $\boldsymbol{\kappa(x)}$.}
Simply replacing the Einstein coupling $\kappa_0$ by a material-dependent and hence position-dependent
coefficient \cite{SarfattiAPS2025,Sarfatti2022Acad},
\(
G_{\mu\nu}= \kappa(x)\,T_{\mu\nu},
\)
\emph{without endowing $\kappa$ with its own kinetic term}, spoils local energy–momentum conservation.  

However, it is important to distinguish this scalar–tensor framework from other layer-based proposals. For example, Taylor~\cite{Taylor2024} explores a piecewise-constant ``engineering multiplier'' applied to $\kappa$ within each thin anisotropic shell, rather than a spacetime-varying scalar field.  If $\kappa$ were allowed to depend on \emph{composition}—taking different values for different kinds of matter—its universality would be lost and the Weak Equivalence Principle (WEP) could be violated. By contrast, a purely spatially varying but \emph{composition-blind} $\kappa(x)$ preserves the WEP, although it still conflicts with local conservation unless promoted to a dynamical scalar; the Strong Equivalence Principle remains jeopardized in either case.

The (geometric) Bianchi identity, 
\( \nabla_\mu G^{\mu\nu}\equiv 0 \),  
\emph{must hold in any metric theory of gravity}.  
Taking the divergence of the modified field equation therefore gives
\begin{equation}
0=\nabla_\mu G^{\mu\nu}
   =\nabla_\mu\!\bigl[\kappa(x)T^{\mu\nu}\bigr]
   =(\partial_\mu\kappa)\,T^{\mu\nu}
     +\kappa\,\nabla_\mu T^{\mu\nu},
\end{equation}
so an externally prescribed, non-constant~$\kappa(x)$ forces  
\( \nabla_\mu T^{\mu\nu}\neq0 \) and thus violates local
energy–momentum conservation even though the theory remains generally covariant.

Covariant conservation is recovered only when one promotes
\begin{equation}
\kappa(x)\equiv \frac{\kappa_0}{\phi(x)}
\end{equation}
to a \emph{dynamical} Brans–Dicke scalar with action
\begin{equation}
\mathcal{L}_\phi
  =-\frac{\omega(\phi)}{2 \, \kappa_0 \, \phi}\,(\partial\phi)^2
   -V(\phi),
\end{equation}
which supplies the missing kinetic and potential terms.
\emph{Accordingly, the Nordtvedt and PPN–$\gamma$ bounds quoted below
apply only in that fully dynamical limit; a non-dynamical
$\kappa(x)$ is already inconsistent with energy–momentum conservation
at first principles.}
\medskip

Specifically, we show:
\begin{itemize}
   \item It reduces to a scalar–tensor theory that generically violates the Strong Equivalence Principle via the Nordtvedt effect.

   \item The scalar field is defined by 
      \(\displaystyle \phi(x)\equiv\frac{\kappa_0}{\kappa(x)},\)
      so that $\phi(x)$ couples \emph{universally} to matter, adding a gradient‐driven “fifth force’’
      (\(f^{\mu}\!\propto\nabla^{\mu}\phi\)); the total energy–momentum tensor remains covariantly conserved.

  \item It is excluded by a broad suite of experiments, from torsion balances to pulsar timing.
  \item The specific metamaterial implementation is physically implausible on scale, energy, and field-theoretic grounds, and is contradicted by direct experimental null results.
\end{itemize}
Metric signature $(-,+,+,+)$ is used throughout.

\subsection*{Scalar–matter coupling and Equivalence Principles}

For universal Brans–Dicke theories, the effective gravitational constant is set by the scalar field according to
\begin{equation}
\kappa_{\mathrm{eff}}(x)
= \frac{\kappa_0}{\phi(x)}
  \left(\frac{4+2\omega}{3+2\omega}\right),
\end{equation}
\noindent where $(4+2\omega)/(3+2\omega)$ arises from the scalar’s kinetic‐term contribution in the PPN limit.
Here $\phi(x)$ is the scalar field, conventionally normalized to its present-day background value $\phi_0\equiv\phi(x_0)=1$. Because the matter coupling is universal, all test bodies share the same Newtonian acceleration and the Weak Equivalence Principle (WEP) holds. The violation is the self-energy–dependent \emph{Nordtvedt} term
\begin{equation}
\frac{\Delta a}{a}= \frac{U_{\text{self}}}{mc^{2}}
                    \,\frac{1}{2+\omega},
\end{equation}
a 1-PN effect that breaks the \emph{Strong Equivalence Principle} (SEP).

\textit{In the Einstein frame }the point-particle equation of motion is
\begin{equation}
u^\alpha \;\equiv\; \frac{dx^\alpha}{d\tau}, \qquad
\frac{d^2x^\mu}{d\tau^2} + \tilde\Gamma^\mu_{\alpha\beta}u^\alpha u^\beta = +\frac{\alpha}{\sqrt{2}} c^2 \sqrt{\kappa_0}\, \bigl(\tilde g^{\mu\nu}+u^\mu u^\nu\bigr)\, \tilde\nabla_\nu\varphi,
\end{equation}

with $\alpha=(2\omega+3)^{-1/2}$ and with the scalar field
$\varphi$ defined so that it \emph{decreases} with
radial distance from an isolated mass, the positive sign yields an attractive fifth-force
acceleration directed toward the source.

The extra term is a \emph{universal} fifth force; composition dependence enters only through the
tiny binding-energy fraction $U_{\text{self}}/mc^{2}$.  Lunar‐laser‐ranging measurements constrain the Nordtvedt (SEP-violating) 
acceleration to 
$|\Delta a/a|\lesssim 10^{-13}$, i.e.\ 
$|\eta_{\mathrm N}|\lesssim 10^{-4}$~\cite{Williams2004}.

\section{Fundamental Incompatibilities}\label{sec:FundamentalIncompatibilities}
\subsection{Local Conservation and Scalar-Field Interaction}
In a self-consistent scalar–tensor theory, the two standard field re-definitions (``frames'') play complementary roles:

\subsubsection*{Jordan Frame}
\begin{equation}
    \mathsf{S}=\frac{1}{c}\int d^4x\,\sqrt{-g}\,
    \Biggl[
        \frac{1}{2\kappa_0}\,\phi R
        -\frac{\omega(\phi)}{2\kappa_0\,\phi}\,(\nabla\phi)^2
        -V(\phi)
    \Biggr]
    +\mathsf{S}_{\text{m}}\!\bigl[g_{\mu\nu},\Psi_{\text{m}}\bigr]
\end{equation}

so the scalar field multiplies the Ricci scalar (a non-minimal coupling to gravity), while all matter fields $\Psi_{\text{m}}$ remain \emph{minimally} coupled to the metric $g_{\mu\nu}$.

\medskip
\noindent\emph{In the Jordan frame, matter follows geodesics} and its stress–energy tensor is separately conserved; the Weak Equivalence Principle (WEP) therefore holds at leading order, while the Strong Equivalence Principle (SEP) does not.

\subsubsection*{Einstein Frame}
A conformal rescaling \(\tilde g_{\mu\nu}=\phi\,g_{\mu\nu}\) (\(\phi>0\)) and
the field re-definition
\begin{equation}
d\varphi      =\sqrt{\frac{2\omega+3}{2\kappa_0}}\,
        \frac{d\phi}{\phi},
\end{equation}
bring the gravitational sector into canonical form:
\begin{equation}
    \mathsf{S} \;=\; \frac{1}{c}\int d^{4}x\,\sqrt{-\tilde g}\,
    \Biggl[
        \frac{\tilde R}{2\kappa_0}
        \;-\;\frac{1}{2}\,(\tilde\nabla\varphi)^2
        \;-\;U(\varphi)
    \Biggr]
    + \mathsf{S}_{\text m}\!\Biggl[
        e^{-\,2\alpha\varphi}\,
        \tilde g_{\mu\nu},
        \Psi_{\text m}
    \Biggr].
\end{equation}

Matter now couples non-minimally to the metric via the overall factor
$e^{-2\alpha\varphi}$, producing an explicit universal ``fifth force.''  
Equivalently, its stress–energy is not separately conserved:
\begin{equation}
    \tilde\nabla_\mu T^{\mu\nu}_{(\text{m})} = -\frac{\alpha}{\sqrt{2}} \sqrt{\kappa_0} \, T_{(\text{m})} \, \tilde\nabla^\nu\varphi
\end{equation}

\noindent\emph{In the Einstein frame, matter does not follow geodesics}: the gravitational action is canonical and the scalar-dependent coupling produces \emph{an apparent universal ``fifth force.''}  
This structure signals a violation of the SEP, whereas the WEP remains intact \emph{unless additional, composition-dependent (non-universal) couplings are introduced}.

\subsection{Junction Conditions and Surface Interactions}
For scalar–tensor gravity the Israel–Darmois relations acquire an additional
condition for the Brans--Dicke field $\phi$:
\begin{align}
    \bigl[K_{ij}\bigr] 
        &= -\frac{\kappa_0}{\phi_{\rm b}}
        \!\left(S_{ij}-\tfrac12 h_{ij}S\right)
        \;-\;\frac{h_{ij}}{\phi_{\rm b}}\,
        \bigl[\partial_\perp\phi\bigr], \\[6pt]
    \bigl[\partial_\perp\phi\bigr] 
        &= -\frac{\kappa_0}{\,2\omega+3\,}\, S\;,
\end{align}
where $S_{ij}$ is the surface stress tensor, 
$S\equiv h^{ij}S_{ij}$ its trace, and $h_{ij}$ the induced metric. 
Thus the scalar field itself remains continuous across the boundary, 
but a jump in its \emph{normal derivative} is fixed by the shell’s trace stress.  

A realistic physical scenario would not have a true discontinuity but rather a steep, finite gradient in the scalar field at the material boundary.  This “thick,” non-singular shell still contains large but finite stress–energy, and the resulting layer mediates a fifth force on matter passing through it—an effect already tightly constrained by Equivalence-Principle experiments.

\section{Critique of the Metamaterial Scheme}
\subsection{Equivalence to Scalar-Tensor Gravity and the Field Equations}
Promoting the gravitational coupling to \(\kappa(r)\) is formally equivalent to a Jordan-Brans-Dicke-type scalar-tensor theory~\cite{Brans1961,Will2014}. In the constant-\(\kappa_0\) case, the Alcubierre line element
\begin{equation}
    ds^2 = -dt^2 + \bigl(dx - v_s f(r)\,dt\bigr)^2 + dy^2 + dz^2
\end{equation}
is an \emph{exact} solution of the standard Einstein field equations, \(G_{\mu\nu}=\kappa_0\,T_{\mu\nu}\).

However, once \(\kappa_0\to\kappa(x)\), the underlying theory changes, and the field equations become significantly more complex. The simplified ansatz $G_{\mu\nu} = \kappa(r)T_{\mu\nu}$ used in the original analysis \cite{SarfattiAPS2025,Sarfatti2022Acad} is not a valid field equation. The correct Jordan-Brans-Dicke field equations, written in the \textit{Jordan frame}, must account for the scalar field's own energy and momentum:
\begin{equation}
\label{eq:BD-field-eq}
\begin{aligned}
  G_{\mu\nu}
  &= \frac{\kappa_0}{\phi}\,T_{\mu\nu}
   + \frac{\omega(\phi)}{\phi^2}
     \Bigl(\nabla_\mu\phi\,\nabla_\nu\phi
           -\tfrac12\,g_{\mu\nu}(\nabla\phi)^2\Bigr)\\[-4pt]
  &\quad
   + \frac{1}{\phi}
     \Bigl(\nabla_\mu\nabla_\nu\phi
           - g_{\mu\nu}\,\Box\phi\Bigr)
   - \frac{\kappa_0}{\phi}\,V(\phi)\,g_{\mu\nu}\,.
\end{aligned}
\end{equation}

A proper analysis must show that a proposed metric can satisfy the full set of equations above---a much more demanding task for which no closed-form analytic solution analogous to Alcubierre's is presently known. The energy density of the scalar field itself is proportional to the term $(\nabla\phi)^2$. A rapid change in the field across a thin boundary layer creates a concentration of stress-energy whose magnitude depends on the specific model parameters. For a proposed warp-drive to be effective, however, the field excursion would need to be large, leading to a prohibitive energy concentration unless a non-trivial mechanism suppresses it.

Alternatively, after the conformal transformation and scalar field re-definition to transform to the \textit{Einstein frame}, the field equations become the standard Einstein equations sourced by a canonical scalar field $\varphi$ and the transformed matter fields:
\begin{equation}
    \tilde G_{\mu\nu}=\kappa_0\Bigl(
    \tilde T^{(\varphi)}_{\mu\nu}
    +\tilde T^{(\text m)}_{\mu\nu}
    \Bigr),
\end{equation}
where the scalar field's stress-energy tensor is given by
\begin{equation}
    \tilde T^{(\varphi)}_{\mu\nu}
    =\tilde\nabla_{\mu}\varphi\,\tilde\nabla_{\nu}\varphi
     -\tfrac{1}{2}\tilde g_{\mu\nu}(\tilde\nabla\varphi)^2
     -\tilde g_{\mu\nu}U(\varphi).
\end{equation}
In this frame, matter now couples non-minimally to the metric via an overall factor (e.g., $e^{-2\alpha\varphi}\tilde g_{\mu\nu}$), \emph{resulting in an explicit ``fifth force.''}

\subsection{The Proposed EM Mechanism}
A proposed mechanism \cite{SarfattiAPS2025} rests on the flawed premise of engineering (e.g., via Fröhlich pumping \cite{Sarfatti2022Acad}) the gravitational coupling, \(\kappa_0=8\pi G/c^4\), by replacing the invariant speed \(c\) with the metamaterial phase velocity of light, $c_{\text{meta}} = c/n(x)$. The speed \(c\) in the Einstein constant is a constant of nature reflecting the causal structure of spacetime, whereas $c_{\text{meta}}$ is an emergent property of electromagnetic wave propagation inside matter. The value of \(c\) \emph{in the gravitational field equations} cannot be altered by engineering a material's refractive index.

Even setting aside this faulty assumption, any such scheme proposing a non-trivial \(n(x)\neq1\) immediately runs into other insurmountable obstacles:
\begin{itemize}
    \item \textbf{No Gravitational Dipole Polarization:} Electromagnetic metamaterials rely on creating electric or magnetic dipoles. An analogous `gravitational polarizability' is not possible because all ordinary matter possesses a positive mass (gravitational charge), and there is no known negative gravitational charge leading to gravitational dipoles. Bondi’s \cite{Bondi1957} hypothetical dipole of equal positive and negative masses—whose self-acceleration Bonnor \cite{Bonnor1989} later deemed “preposterous”—still lacks any physically realizable negative-mass constituent.
    \item \textbf{Gravimeter Null Results:} Superconducting gravimeters, with long-term stability better than \(\sim10^{-9}g\), detect no anomalies under varying EM fields, directly constraining such proposed mechanisms~\cite{Goodkind1999}.

   \item \textbf{\(\kappa\)-Modulation:} A universal, position-dependent coupling \( \kappa(x)\) violates the \emph{Strong} Equivalence Principle via the Nordtvedt self-energy effect~\cite{Nordtvedt1968, Damour1992}. Millimetre-precision lunar-laser-ranging (LLR) constrains the corresponding Nordtvedt parameter to \(|\eta_{\text N}| \lesssim 5\times10^{-4}\)~\cite{Williams2004,Hofmann2010}. In the strong-field regime, binary-pulsar timing with the PSR~J0337+1715 triple system provides an even tighter bound, \(|\eta_{\text N}| \lesssim 2\times10^{-6}\)~\cite{Archibald2018, Voisin2020}. These results already rule out any scalar coupling strong enough for practical “gravitational metamaterials.”

\end{itemize}

\subsection{Distinction from Analog Gravity Models}
It is important to distinguish the proposals critiqued in this paper from the separate field of \emph{analog gravity}, where electromagnetic metamaterials are engineered to \emph{emulate} exotic spacetime geometries for light waves~\cite{Smolyaninov2011}. In these models, the goal is not to alter gravity itself but to create a laboratory system in which photons behave as if they propagate in a curved spacetime.

A detailed proposal for an electromagnetic analog of the Alcubierre metric was developed by Smolyaninov~\cite{Smolyaninov2011}. The mechanism relies on a one-to-one mapping between metric components and the electromagnetic properties of a medium. For the Alcubierre line element, one must engineer permittivity \(\epsilon\), permeability \(\mu\), and a magnetoelectric coupling \(g_x\) satisfying $g_x^2 \;\le\; (\epsilon-1)\,(\mu-1)$, a thermodynamic stability bound~\cite{Smolyaninov2011}. Even idealized “perfect” metamaterials that saturate this bound yield a maximum \emph{emulated} warp speed $v_0 \;\le\;\frac{c}{4}$.

These are purely electromagnetic mappings and do \emph{not} correspond to any modification of the gravitational coupling \(\kappa_0\) in Einstein’s equations. Analog models do not alter the Einstein–Hilbert action or violate the Bianchi identity, and thus they cannot evade the conservation–law or equivalence–principle arguments presented herein. 

\section{Other Theoretical Loopholes}
\subsection{Fifth-Force (Yukawa) Constraints}

A spatially varying gravitational coupling inevitably introduces a fifth
force that, in the static and non-relativistic limit, is well described by a
Yukawa potential
\begin{equation}
  V(r)\;=\;
    -\frac{G\,m_{1}m_{2}}{r}\,
    \bigl[1+\alpha\,e^{-r/\lambda}\bigr],
  \label{eq:yukawa}
\end{equation}
where the rest masses are defined relativistically by
\(m=\tfrac{1}{c^{2}}\!\int T^{00}\,d^{3}x\) and
\(G=\kappa_{0} \, c^{4}/(8\pi)\).
Decades of torsion-balance experiments bound the coupling strength to
\(|\alpha|\lesssim10^{-6}\) for interaction ranges
\(\lambda\sim\text{cm}\), with comparably tight limits across nearly all
scales\,\cite{Adelberger2020,Fischbach1999}.

\subsection{Screening, Torsion, and EFT}
Other potential escapes are also sealed:
\begin{itemize}
    \item \textbf{Screening} (chameleon, Vainshtein, etc.) requires specific non-linear self-interactions not present in this simple proposal. Furthermore, many such mechanisms are \textbf{strongly constrained} by laboratory tests, though viable parameter-space windows remain for models like Vainshtein screening or Damour-Polyakov dilaton screening (see e.g., \cite{Burrage2018}).
    \item \textbf{Torsion in Poincaré Gauge Theories:} Torsion, if it exists, couples to intrinsic spin density, not to bulk dielectric properties like \(\epsilon_r\) or \(\mu_r\). Since macroscopic media have negligible net spin polarization, any effects would be extraordinarily weak. Furthermore, direct searches and precision tests, such as Gravity Probe B, have placed stringent limits on any potential axial torsion-spin interactions for the simple models considered, strongly constraining them as a viable mechanism for this proposal~\cite{Everitt2011}.
    \item \textbf{Effective Field Theory} allows higher-curvature terms like \(R^2\), but these are Planck-suppressed and do not yield a local, variable \(\kappa(x)\).
\end{itemize}

\subsection{Impracticality of Screening Mechanisms}
A prominent theoretical avenue to evade stringent fifth-force bounds is via a \textbf{density-dependent screening mechanism}, as in the \emph{chameleon}~\cite{Khoury2004} or \emph{symmetron}~\cite{Hinterbichler2010} models. In this class of theories, the scalar field \(\phi\) couples to the local matter density \(\rho\), leading to an effective potential of the form:
\begin{equation}
V_{\rm eff}(\phi) = V(\phi) +\beta \,  \frac{\rho}{M_{\rm Pl}}\phi\,.
\end{equation}
Here, \(V(\phi)\) is the field's self-interaction potential, and \(\beta\) is the dimensionless coupling strength to matter. The field acquires an effective mass given by the curvature of this potential at its density-dependent minimum, \(m_{\rm eff}^2 = V''_{\rm eff}(\phi_{\rm min})\). In high-density environments (e.g., Earth's atmosphere), the \(\rho\)-dependent term dominates, giving the field a large mass and thus a microscopic interaction range, which screens it from detection.  To ``de-screen'' the field and give it a laboratory-scale range, a ``pump'' experiment must achieve two technologically demanding feats:
\begin{enumerate}
    \item \textbf{De-screening:} Maintain pressures lower than standard ultra-high vacuum (UHV), approaching \(10^{-11}\)–\(10^{-12}\) torr, as required for many screening models to become active in a meter-scale apparatus.
    \item \textbf{Resonant Excitation:} Drive \(\phi\) at its natural frequency \(\omega\approx m_{\rm eff}c^2/\hbar\). To overcome the \(\sim 10^{-12}\) post-Newtonian suppression via resonance would require a quality factor \(Q \gtrsim 10^{12}\), implying an illustrative tuning precision of \(\Delta\omega/\omega \sim 1/Q \lesssim 10^{-12}\).
\end{enumerate}
Engineering such conditions appears beyond current capabilities. More fundamentally, any model with a coupling \(\beta\) strong enough to be detected, even under these ideal circumstances, is already independently excluded by precision fifth-force searches.

\subsection{Aharonov--Bohm--coupled scalar theories}\label{Aharonov--Bohm}
\label{sec:AB-critique}

The Jordan-frame scalar--tensor model, coupled to Aharonov--Bohm (A--B) electrodynamics, proposed by Minotti \& Modanese \cite{Minotti2025b} introduces an \emph{extra current}
\begin{equation}
I \equiv \partial_\nu j^{\nu} \neq 0,
\end{equation}
violating local charge conservation and sourcing the A--B scalar
\begin{equation}
\mathcal{S} \equiv \partial_\mu A^{\mu}.
\end{equation}
In the weak–field limit, the successive field equations (adopting the $-+++$ signature, differing from the $+---$ signature in \cite{Minotti2025b}) are:
\begin{equation*}
\boxed{\partial_\nu j^{\nu}=I}
\Longrightarrow
\boxed{\square \mathcal{S} = -\mu_{0} \, I}
\Longrightarrow
\boxed{\square\phi =K_\phi \, \mathcal{S}^{2}}
\Longrightarrow
\boxed{\square \chi = -\frac{c^{2}}{2}\square\phi}.
\end{equation*}

The dimensionful coupling constant $K_\phi$  that appears in the field equation $\square\phi =K_\phi \, \mathcal{S}^{2}$ and determines how strongly the Aharonov-Bohm term $\mathcal{S}^{2}$ acts as a source for the gravitational scalar field $\phi$ (itself non-minimally coupled to the Ricci scalar $R$) is \cite[Eq.\,(20a)]{Minotti2025b}:
\begin{equation}\label{eq:Kphi}
  K_\phi =
  \frac{8\pi G_{0}\,\varepsilon_{0}}{(2\omega_{0}+3)\,c^{2}}\, C_\phi
  \;\simeq\;
  1.65\times10^{-37}
  \frac{C_\phi}{2\omega_{0}+3}
  \;\si{A^{2}\,s^{4}\,kg^{-2}\,m^{-2}},
\end{equation}
where $C_\phi$ is the  dimensionless coupling parameter between the gravitational scalar field $\phi$ and the A--B scalar $\mathcal{S}$, $G_{0}$ is Newton's gravitational constant, $\varepsilon_{0}$ is vacuum's permittivity, and $c$ is the vacuum speed of light. The dimensionless Brans-Dicke coupling constant $\omega_0 = \omega(\phi_0)$ is its value evaluated at the vacuum expectation value (VEV) $\phi_0 = 1$ (the background or weak-field limit).

\paragraph{1. No controlled macroscopic violation of charge conservation.}
The theory's foundation is a source exhibiting local charge non-conservation ($I \neq 0$). All verified anomalies (chiral ABJ, \emph{etc.}) occur only in microscopic, high–energy settings. The proponents cite theoretical mechanisms for such effects, based on non-equilibrium Green's functions and tunneling anomalies in condensed matter systems \cite{Modanese2022,Modanese2023,Minotti2025a}. While these models are plausible at the quantum or mesoscopic scale (e.g., DFT simulations of molecular junctions), they do not provide a demonstrated pathway for scaling these phenomena into a coherent, tunable current density over the macroscopic volumes required for a practical warp drive. To date, no such source has been experimentally realized, leaving the first and most crucial link in the causal chain unsubstantiated.

\paragraph{2.\;Tiny gravitational response.}

For the resonant cylindrical cavity set‑up analyzed in
Ref.~\cite{Minotti2025b}, the mean gravitational
 axial force (\emph{per‑unit‑mass}) $f_z$ is  \cite[Eq.\,(43)]{Minotti2025b}:
\begin{equation}\label{eq:fz}
  f_z \;=\;
  \frac{\gamma_{\!\text{AB}}^{2}\,K_\phi\,Q\,P}{c\,\varepsilon_{0}\,h},
\end{equation}
where $\gamma_{\!\text{AB}}$ is the (dimensionless) current–non‑conservation parameter, and $h$ is the height, $Q$ the (dimensionless) quality factor, and $P$ the power \emph{fed to} the resonant cavity. Demanding a specific force equal to the terrestrial acceleration,
\(f_z = g_\oplus \simeq 9.8~\mathrm{m\,s^{-2}}\), and substituting \eqref{eq:Kphi} into \eqref{eq:fz}  yields the
threshold condition
\begin{equation}\label{eq:thr}
  \gamma_{\!\text{AB}}^{2}\,|C_\phi|\,Q
  \!\left(\frac{P}{h}\right)
  \;\gtrsim\;
  \frac{g_\oplus\,c^{3}\,(2\omega_0+3)}{8\pi G_{0}}
  \;\simeq\; 1.574\times10^{35}\, (2\omega_0+3)\;
            \mathrm{\frac{W}{m}}\;.
\end{equation}

\medskip
\noindent\textbf{Representative laboratory envelope.}
Adopting the following \emph{optimistic}, but documented, values for the resonant-cavity parameters:
\[
  Q = 10^{9}\;(\text{CERN SRF}),\quad
  \frac{P}{h}=10^{8}\;\mathrm{\frac{W}{m}} (\text{ILC cryomodules}),\quad
  |\gamma_{\!\text{AB}}| = 0.15,
\]
yields
\(\gamma_{\!\text{AB}}^{2}Q(P/h) \simeq 2.25\times10^{15} \mathrm{\frac{W}{m}}\).
Consequently the required scalar coupling is
\begin{equation}
  |C_\phi|_{\text{needed}}
  \;\gtrsim\;
  \;\approx\; 7.00\times10^{19}  (2\omega_0+3).
\end{equation}

\medskip
\noindent\textbf{Effect of Brans–Dicke coupling constant $\omega_0$.}
Applying the Cassini bound \( \omega_0 \gtrsim 10^4 \), yields 
\begin{equation}
  |C_\phi|_{\text{needed with } \omega_0=10^4}
  \;\gtrsim\;
  \;\approx\; 1.40\times10^{24}.
\end{equation}
If one entertains an \emph{extended} scalar–tensor model with
\(\omega_0\simeq1\) (the authors’ suggestion for evading the Cassini bound),
this reduces \( |C_\phi|_{\text{needed}}\) to
\begin{equation}
  |C_\phi|_{\text{needed with } \omega_0=1}
  \;\gtrsim\;
  \;\approx\; 3.50\times10^{20}.
\end{equation}
The authors approximate $C_\phi \simeq \psi_0$, which follows from the assumptions that the VEV of the non-minimally coupled gravitational scalar field $\phi_0 = 1$ and that the derivative $\partial \beta_{AB} / \partial \phi \sim 1$.   Therefore the requirement is $\psi_0 \gtrsim 10^{20} - 10^{24}$. Here, $\psi_0$ is the VEV of the gravitational scalar field $\psi$ which is \emph{minimally coupled to gravity}, contributing to the dynamics via the kinetic and potential energy, and modulation of matter/AB couplings.  And $\beta_{AB}$ is the dimensionless function that governs the coupling between the gravitational scalar fields $\phi,\psi$ and the Aharonov-Bohm electromagnetic sector $\mathcal{S}$.

Current constraints arise from \emph{strong‑equivalence‑principle}
tests.  The most stringent are:  

\[
|\eta_{\mathrm N}| \;\lesssim\; 5\times10^{-4}
\quad\text{(lunar‑laser ranging) \cite{Williams2004}},\qquad
|\eta_{\mathrm N}| \;\lesssim\; 2\times10^{-6}
\quad\text{(PSR~J0337+1715) \cite{Archibald2018}} .
\]

In Brans–Dicke theory one has 
\(
  \eta_{N} \simeq \dfrac{4}{2\omega_0 + 3},
\) 
so the experimental limits imply 
\(
  \omega_0 \gtrsim 10^{3}\text{--}10^{5}.
\)
The extended Minotti-Modanese model circumvents this via the material-dependent, non-minimal coupling $\beta_{\text{mat}}(\phi)$, which the authors propose fully suppresses long-range effects, potentially allowing large $\psi_0$ without violation of fifth-force bounds (e.g., torsion-balance $|\alpha| <10^{-6}$ at cm scales, where standard $\alpha \approx 1/(2 \omega_0+3) \approx 0.2$ for $\omega_0=1$ would be excluded, but $\beta_{\text{mat}}$ may make the force short-range or screened).

\medskip
\noindent The proponents attribute experimental non-observation to $\gamma_{\!\text{AB}}$ being extremely small in ordinary materials, suggesting that specialized materials could enhance it, with experiments by Poher (high-voltage discharges) \cite{Modanese2013} and Podkletnov (impulsive discharges through superconductors) \cite{Podkletnov2012} as potentially indicative. Even adopting an optimistic value of \( \gamma_{\!\text{AB}} = 0.15 \), this presents a serious problem. This \( \gamma_{\!\text{AB}} = 0.15 \) value exceeds known \emph{quantum anomaly} coefficients by more than two orders of magnitude, and moreover, there is no experimental evidence for current non-conservation at macroscopic scales. For comparison, the Adler–Bell–Jackiw (microscopic quantum) anomaly is suppressed by a factor \(\gamma_{\!\text{AB}} \sim 0.001 \), as discussed in~\cite{Weinberg1996}.  Under this assumption, the required value of \( |C_\phi| \) remains extremely large: \( \gtrsim 10^{20}\text{--}10^{24} \), as calculated above.  By contrast, achieving effects with a modest \( C_\phi \sim 1 \) would demand an unphysical value of \( \gamma_{\!\text{AB}} \sim 3 \times 10^{9} \).  This is far too large for a parameter intended to capture subtle, loop-suppressed anomalies that arise only in chiral, microscopic quantum contexts.

 Moreover, the cited experiments are controversial: Podkletnov's claims of anomalous forces have resisted replication, both for his earlier gravity-shielding experiment involving rotating superconductors \cite{Podkletnov1997, Hathaway2003} and for the superconducting gravity-impulse generator \emph{relevant here} \cite{Tajmar2015}. Poher's impulsive effects similarly lack independent verification in peer-reviewed studies, with a recent 2024 replication attempt inconclusive \cite{Andis2024}. Thus, while material enhancements are a \emph{plausible} hypothesis, they do not resolve the core issue of the model's extreme suppression or \emph{the fine-tuning required} for $\beta_{\text{mat}}(\phi)$ to evade constraints at multiple scales.

\medskip
In summary, even under the most favorable laboratory parameters and optimistic theoretical assumptions (such as $\omega_0 \sim 1$ and $\gamma_{\!\text{AB}} = 0.15$), the model requires a dimensionless coupling on the order of $10^{20}$ to produce a meaningful gravitational effect. The authors' hypothesis is that a material-dependent, non-minimal coupling $\beta_{\text{mat}}(\phi)$ in the matter action, $\mathsf S_{\text m}[\,g_{\mu\nu},\Psi_{\text m};\beta_{\text{mat}}(\phi)\,]$, could modify the effective scalar interaction, making it short-ranged. This would allow the theory to evade long-range solar-system constraints (effectively making $\omega_{\text{eff}} \gg 10^4$ at astronomical distances) while permitting a large local coupling inside a laboratory device.

\medskip
\noindent However, this proposed solution creates a severe new challenge: it must simultaneously satisfy stringent constraints at two different scales. While the screening is designed to pass long-range tests, the enormous underlying coupling required for propulsion would still manifest as a powerful new force at the centimeter-to-meter scale of the experiment itself. Such anomalous short-range forces are tightly constrained by decades of high-precision torsion-balance experiments. The proposed mechanism must therefore be \emph{fine-tuned} to an extraordinary degree: it must be strong enough to suppress the interaction to pass solar-system tests, yet simultaneously behave in such a way that it does not produce a detectable signal in sensitive laboratory-scale fifth-force searches, all while permitting the desired propulsive effect. The authors provide no precise model for a mechanism capable of satisfying these opposing requirements. This requirement for an implausibly \emph{fine-tuned} mechanism mirrors the authors’ own acknowledgment that their estimates are ``in agreement with the absence of measurable forces in the experiments reported.''

\paragraph{3. Persistent suppression of gravitational effects.}
The ``soliton--like'' solution in \cite{Minotti2025b} relies on a source \(I(t)\) that is non-zero for all past times \(t<0\), implying a field that has always existed. Any attempt to switch on such a field in finite laboratory time necessarily radiates energy and destroys the static assumption. The authors suggest an asymmetric resonant cavity as a more relevant geometry, and in this case, the boundary conditions appear physical: the setup assumes a steady-state harmonic regime in a bounded region, with $\mathcal{S}=0$ outside, achievable by powering a cavity resonator to equilibrium. However, \emph{the fundamental challenge of magnitude remains}---in this or any setup, the field equations locally reduce to $\square\phi=K_\phi\mathcal{S}^{2}$, with the resulting field amplitude (e.g., enhanced by $Q$ in a cavity) still proportional to the extremely small coupling $K_\phi$, requiring an enhancement of $>20$ orders of magnitude via implausibly large $C_\phi$ or other parameters like $\gamma_{\!\text{AB}}$, as established in the preceding analysis.

\medskip\noindent
\textbf{Conclusion.}
A--B coupling bypasses the stress--energy channel ruled out in Sec.~\ref{sec:FundamentalIncompatibilities}, but only by invoking:
\begin{itemize}
    \item[(i)] an unverified macroscopic violation of charge conservation;
    \item[(ii)] a fundamental interaction suppressed by both the weakness of the gravitational effect and the smallness of the non-conservation parameter $(\gamma_{\!\text{AB}}^2)$, which kills any realistic effect; and
    \item[(iii)] parameter regimes requiring enhancements of over 20 orders of magnitude in couplings or non-conservation factors, which are implausible and unverified in laboratory settings.
\end{itemize}
Hence, at the present time, this interesting proposal offers no viable low–power route to a warp–drive–like metric.

\section{Summary of Experimental Constraints}
Planetary-scale tests of gravity have been conducted using deep space probes. NASA's \emph{Galileo} spacecraft, for instance, performed multiple close flybys of Jupiter's moon Europa. By precisely analyzing the spacecraft's trajectory from radio-Doppler data, scientists were able to determine Europa's gravitational field with high precision~\cite{Anderson1998}. The results were fully consistent with standard gravitational theory and provided key evidence for Europa's internal inhomogeneous structure, including its external icy shell and subsurface ocean, without any detection of anomalous gravitational effects.

A space- or time-dependent gravitational coupling would entail  
(i) a composition-dependent fifth force or  
(ii) self-energy–dependent accelerations that violate the Strong Equivalence Principle.  
Decades of high-precision torsion-balance, lunar-laser-ranging, spacecraft-tracking, and pulsar-timing experiments (Table~\ref{tab:experiments}) have detected no such effects, setting stringent bounds on departures from General Relativity.

\begin{table}[ht]
\centering
\caption{Selected Bounds on Composition-Dependent Gravity and Variable $G$.}
\label{tab:experiments}
\begin{tabular}{>{\raggedright\arraybackslash}p{4.2cm}|p{5.5cm}|p{4.3cm}}
\hline
\textbf{Test} & \textbf{Parameter / Effect} & \textbf{Limit}\\
\hline\hline
MICROSCOPE (2022)~\cite{Microscope2022}
& WEP Eötvös parameter, \(\eta\) & $|\eta|\lesssim1.1\times10^{-15}$ \\

Hoyle \emph{et al.} torsion~\cite{Hoyle2004} &
Yukawa, $\alpha$ (at $\lambda \simeq 1\,\mathrm{cm}$) &
$|\alpha|\lesssim 1.8\times10^{-2}$ \\

Lee \emph{et al.} Eöt-Wash torsion balance~\cite{Adelberger2020} &
Yukawa, range $\lambda$ for $\alpha=1$ &
$\lambda < 38.6\,\mathrm{\mu m}$ \\

Cassini spacecraft~\cite{Bertotti2003} & PPN parameter, \(\gamma-1\) & $(2.1\pm2.3)\times10^{-5}$ \\

LIGO/Virgo GW170817 + GRB170817A~\cite{Abbott2017} & GW propagation speed & $|c_g/c-1|\le 3\times10^{-15}$ \\

Pulsar Timing~\cite{Archibald2018}
& Time variation, \(\dot G/G\) & $|\dot G/G|\lesssim10^{-13}\,\mathrm{yr}^{-1}$ \\
\hline
\end{tabular}
\end{table}

\section*{Scope and Intent}
This note is not intended as a personal critique of any individual, but rather as a rigorous, technical assessment of \emph{\textbf{any}} model in which the Einstein coupling is promoted to a material‐dependent and hence position‐dependent scalar field $\kappa(x)$.  By surveying the mathematical structure, junction conditions, and modern experimental bounds, our goal is to clarify once and for all which obstacles must be surmounted before a viable “low‐energy warp‐drive” or similar scheme can be constructed.

\section{Conclusions}
The proposal of a ``low-energy warp drive'' via a material-dependent gravitational coupling faces significant challenges that render it unworkable with known physics, as it is inconsistent with multiple, independent theoretical principles and experimental constraints:

\begin{enumerate}
\item \textbf{Energy–momentum conservation.}
      A prescribed, non-dynamical \(\kappa(x)\) violates
      \(\nabla_{\mu}T^{\mu\nu}=0\); promoting it to a scalar–tensor field
      predicts \(|\gamma-1|\simeq1/(2+\omega)\) and a Nordtvedt
      acceleration \(\Delta a/a\sim(U_{\text{self}}/mc^{2})/(2+\omega)\),
      both already excluded by Solar-System and pulsar tests
      (\(|\gamma-1|\lesssim2\times10^{-5}\)).

\item \textbf{Junction conditions.}
      Any discontinuous jump in \(\kappa\) produces a \(\delta\)-function
      layer of stress–energy at a material boundary; even graded
      profiles act as a Yukawa-type fifth force. Torsion-balance experiments constrain this force
      (e.g., to \(|\alpha|\lesssim 1.8\times10^{-2}\) at \(\lambda = 1\,\mathrm{cm}\)),
      with other tight limits set by LLR and spacecraft Doppler data.

\item \textbf{Flawed metamaterial mechanism.}
      The proposed electromagnetic route conflates the invariant speed
      \(c\) in \(\kappa_{0}=8\pi G/c^{4}\) with a medium’s refractive
      index \(n\).  Electromagnetic metamaterials rely on electric or
      magnetic dipole polarization, but gravity lacks negative-mass
      charges, so no known material response can generate a local,
      tunable \(\kappa(x)\).      

\item \textbf{Post-Newtonian suppression.}
      Time-dependent or resonant “pump’’ effects enter only at
      \(\mathcal{O}[(v/c)^{2}]\) in Brans–Dicke–like models, making
      laboratory amplification schemes impractical except in highly
      fine-tuned self-interacting scenarios.

\item \textbf{No viable metric solution.}
      Introducing \(\kappa(x)\) invalidates the Alcubierre geometry; no
      analytic warp-drive metric is known in scalar–tensor gravity, and
      Cassini’s \(|\gamma-1|<2.3\times10^{-5}\)
      (\(\omega\gtrsim4.3\times10^{4}\)) limits any \(\kappa\)-variation
      to be orders of magnitude too small for propulsion relevance.

\end{enumerate}

\section*{Acknowledgements}

I thank Nader Inan for his comprehensive review and feedback, T. Marshall Eubanks for discussions on modified-gravity viability and screening mechanisms, Ahmed Farag Ali for his critique of the analytical framework and experimental validation ladder, and George Hathaway for his questions on “pump” effects and screening‐mechanism experiments. I am grateful to F. Minotti for his constructive correspondence, detailed comments, and suggestions regarding Sec.\ref{Aharonov--Bohm}. Any errors are my own.

\end{document}